\documentclass[10pt,letterpaper]{article}

\usepackage{fullpage}

\usepackage{amsmath}
\usepackage{amssymb}
\usepackage{graphicx}
\usepackage{hyperref}
\usepackage{mathtools}
\usepackage{authblk}
\usepackage{cite}
\usepackage{dutchcal}
\title{Defect Topology in Colloidal Smectics}
\author[1]{Chaya Halperin}
\author[1]{Hillel Aharoni  \thanks{hillel.aharoni@weizmann.ac.il}$^{,}$}
\affil[1]{Department of Complex Systems, Weizmann Institute of Science, Rehovot 7610001, Israel}
\date{}

\begin{document}
\maketitle

\begin{abstract}
Colloidal smectics -- layered structures formed in dense suspensions of rod-like particles -- often exhibit grain boundaries, across which the layer orientation changes by $90^{\circ}$. Motivated by this feature, we develop a layer-based topological framework that treats orthogonal grain boundaries as constituents of the ground state rather than as exceptional defect structures. Extending the layer-based approach for ordinary smectics, we reduce the smectic structure to layers, half-layers, and domain walls.
We classify the topology of defects and their combination rules based on this structure. In two dimensions, point defects are described by semi-directed cycle graphs. Although the disclination charge remains a valid topological invariant, it does not uniquely classify defects, as distinct graphs may share the same charge. In three dimensions, line defects are classified by their transverse graph structure, while point defects exhibit qualitatively different behavior. In particular, we show that the hedgehog disclination charge is not a topological invariant, but instead varies continuously under smooth deformations of the layer structure.

\end{abstract}

\section*{Introduction}
\begin{figure}
    \centering
    \includegraphics[width=0.8\linewidth]{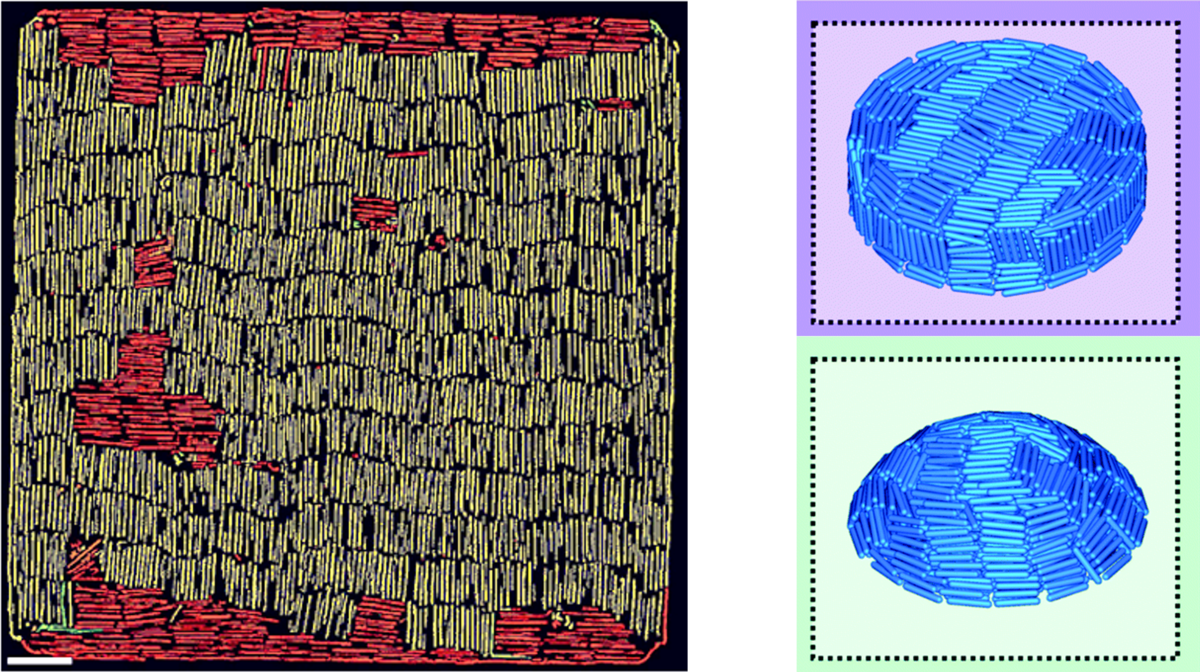}
    \caption{Colloidal smectic textures in confinement, illustrating the occurrence of orthogonal grain boundaries between locally layered domains.
    Left, adapted from \cite{Aarts2017}: Silica rods in a two-dimensional square confinement. Scale bar: $10 \mu m$.
    Right, adapted from \cite{Loewen2022}: Simulation of hard rods confined in three dimensions; cylindrical confinement (top) and spherical confinement (bottom).}
    \label{fig:aarts}
\end{figure}

Colloidal liquid crystals are liquid-crystalline phases formed by suspensions of anisotropic particles, such as rods, whose ordering is governed primarily by excluded-volume interactions. In the hard-rod setting, Onsager \cite{Onsager1949} showed that steric repulsion alone can drive an isotropic--nematic transition, and later works demonstrated that increasing density can further stabilize a smectic phase \cite{frenkel1988thermodynamic, DogicFraden1997, Wensink2013}. The smectic phase combines orientational order with a one-dimensional density modulation, producing a layered structure while remaining liquid-like within each layer. The stability of the smectic phase may depend on subtle geometric nuances, such as the shape of the rod ends \cite{Kamien2023kikibouba}.
Experiments and simulations on colloidal rods have further shown that these smectic states often contain abundant grain boundaries separating neighboring domains whose layer orientation differs by $90^\circ$ \cite{Aarts2017, Wittmann2021particle, Loewen2022, Monderkamp2021, narayan2006nonequilibrium, geigenfeind2015confinement}, see \autoref{fig:aarts}. This ubiquity of orthogonal grain boundaries distinguishes colloidal smectics from the molecular smectic phase, which has a single-domain ground state.

Topological defects are central to the description of ordered phases because they encode invariant properties of the ordered state. In homogeneous phases such as nematics, where the ground state is described by a continuous manifold of equivalent orientations, defects are classified by the homotopy groups of the ground-state manifold \cite{Mermin1979, Kleman1973, Alexander2012, Selinger2024}. Operationally, one surrounds the defect by a measuring loop or sphere and studies the induced map of that measuring manifold into the ground-state manifold; the corresponding homotopy class gives the topological charge. In smectics, however, broken translational symmetry imposes an integrability constraint: not every homotopy class allowed by the director field can be realized by a layered texture. This failure of naive homotopy theory was formalized by Po\'enaru using foliation theory \cite{Poenaru1981}. Building on that perspective, later work characterized the topology of smectic defects using the underlying layer structure rather than the director field alone \cite{ChenAlexanderKamien2009, Machon2019, Aharoni2017}.
 
For colloidal smectics, previous work has employed several order parameters to address $90^\circ$ grain boundaries. In two dimensions (2D), the tetratic order parameter is often used \cite{Wittmann2021particle, Loewen2022, narayan2006nonequilibrium}. By identifying directions related by quarter turns, tetratic order renders orthogonal reorientations of the nematic director continuous. Other works suggested the use of a complex tensor order parameter \cite{Paget2023} to simultaneously capture tetratic order and smectic phase. Analogous constructions have also been explored in three dimensions (3D) \cite{Monderkamp2021}; however, the analogy is limited by the tetratic order's essentially two-dimensional nature. Wittmann \cite{wittmann2024layer} showed that the tetratic endpoint structure of grain boundaries can be recovered from a layer-based description, similar to regular smectics \cite{Machon2019}. These approaches capture important geometric features of colloidal smectic textures, but they do not provide a full layer-based classification of admissible local defect structures.

In this work, we develop a layer-based topological framework for defects in colloidal smectics by extending the regular smectic-layer description to include grain boundaries as part of the ground-state structure. 
More precisely, we treat orthogonal grain boundaries on the same footing as layers and half-layers and use their intersection with measuring circles and spheres to encode the local defect structure. This yields a classification of two-dimensional defects in terms of cyclic graph data and shows that disclination charge, while still meaningful, does not, on its own, classify colloidal smectic defects. We then extend the construction to three dimensions, where line defects are governed by their transverse two-dimensional structure, while point defects require a richer description on the measuring sphere. In particular, we show that the usual hedgehog charge is not a topological invariant of colloidal smectic point defects.

\section*{Theoretical Framework}
We extend the layer-topological framework developed for ordinary smectics by Machon \emph{et al.} \cite{Machon2019}. In this construction, the smectic state is reduced to a \emph{layer structure}, which is composed of codimension-one ``layers'' (density maxima) and ``half-layers'' (density minima). Defects are identified as non-manifold loci on either the layer set or the half-layer set. They are characterized by intersecting the layer structure with a small measuring manifold surrounding the defect. The intersections of layers or half-layers with the measuring manifold encode the local defect topology. In two dimensions, the disclination charge is determined by the number of intersections, denoted by \(m\), through \(q=1-\frac{m}{2}\) \cite{Machon2019, Poenaru1981}. In three dimensions, line defects are classified by their transverse structure: a cross-section perpendicular to the line reduces the problem to a two-dimensional point defect. The transverse charge may change at isolated monopole points along the line, where a cone-shaped layer forms and the charge jumps by an integer \cite{Machon2019, Aharoni2017}. Point defects are characterized differently. Intersecting the layer structure with a measuring sphere yields a graph whose vertices correspond to the connected regions of the sphere and whose edges correspond to the intersection loops. Since the sphere is simply connected, the resulting graph is necessarily a tree. The charge is then determined, up to an overall sign, by the structure of this tree. However, distinct trees may yield the same absolute value of the charge, so the charge alone does not uniquely determine the defect topology \cite{Machon2019}. In that framework, the measuring manifold intersects only layers and half-layers, so the nematic order remains continuous on it. In colloidal smectics, by contrast, the measuring manifold may also intersect domain walls separating orthogonal domains, necessitating an extension of the ordinary smectic classification.

Our key modification in the colloidal case is to include orthogonal grain boundaries in the reference ground state. This is motivated by the observed prevalence of interfaces across which neighboring smectic domains differ by a $90^\circ$ rotation \cite{Aarts2017, Loewen2022, Monderkamp2021}. Accordingly, we consider a layer structure composed not only of layers and half-layers, but also of grain boundaries separating orthogonal domains. In this description, a grain boundary is not treated as an exceptional defect object, but rather as an allowed constituent of the ground-state texture.
The physical motivation for this choice comes from the steric nature of colloidal rod systems. We consider hard rods for which the dominant interactions are excluded-volume interactions: particle overlaps are forbidden, while all non-overlapping configurations are energetically equivalent. Unlike molecular smectics, where intermolecular interactions such as van der Waals forces play an important role, for larger-scale colloidal systems, these energetic interactions are strongly reduced relative to the scale-invariant steric interactions.
Thus, in the large-scale limit, both phase behavior and defect energetics are entropic in essence, arising from the competition between orientational entropy and the reduction of accessible configuration space due to excluded volume \cite{Onsager1949}. Correspondingly, free-energy differences satisfy $\Delta F \approx -T\Delta S$.
These considerations suggest that interfaces between perpendicularly oriented smectic domains may remain comparatively low-cost structures in the steric regime.

\begin{figure}[htb]
    \centering
    \includegraphics[width=0.7\linewidth]{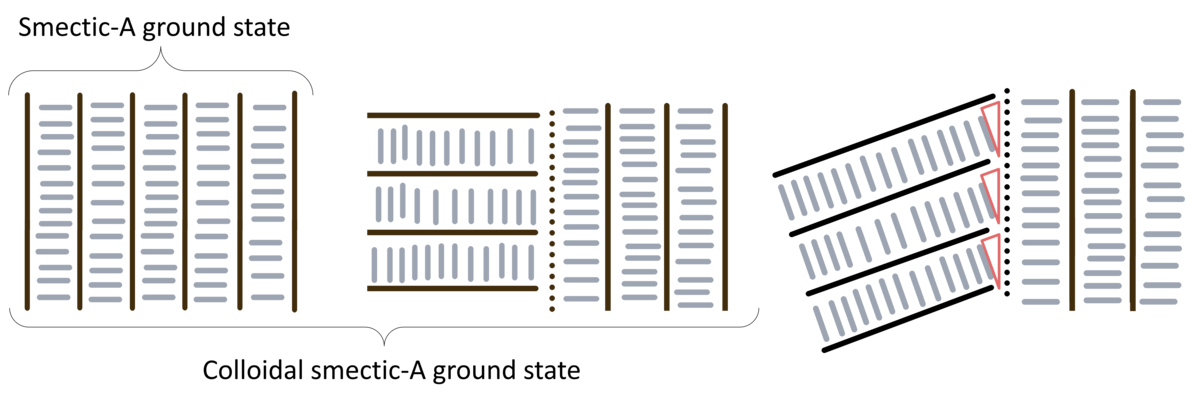}
    \caption{Void formation at an interface between smectic domains. 
    Left: one domain that consist of parallel layers. 
    Center: two domains with mutually orthogonal layers.
    Right: Layers meet an interface at an intermediate angle \(0<\alpha<\pi/2\), creating triangular void regions.
    The void volume or area scales as \(\sin(2\alpha)\), vanishing for parallel and perpendicular alignment.}
    \label{fig:GS}
\end{figure}

To see this, consider a grain boundary and suppose that, on one side, the smectic layers meet the boundary at an angle $0\leq\alpha\leq90^\circ$ (\autoref{fig:GS}). In this geometry, a void region of approximately triangular cross-section appears between neighboring layers. If the edge of the triangle along the boundary has length $\ell$, then the other two sides have lengths $\ell\sin\alpha$ and $\ell\cos\alpha$. Assuming that the boundary extends a distance $L$ in the perpendicular direction, the corresponding void volume is
\(
V(\alpha)
=
\frac{L\ell^2\sin\alpha\cos\alpha}{2}
=
\frac{L\ell^2\sin(2\alpha)}{4}
\). 
The corresponding void volume (area) per unit boundary area (length) is therefore
$\frac{V(\alpha)}{L\ell}
=
\frac{\ell}{4}\sin(2\alpha)$
($ \frac{A(\alpha)}{\ell}
= 
\frac{\ell}{4}\sin(2\alpha)$),
which vanishes for $\alpha\in\{0,\pi/2\}$ and is positive for intermediate angles. Thus, intermediate angles produce a finite excluded-volume penalty in the form of void regions. Consequently, the steric free energy is minimized to approximately zero by orthogonal alignment, with corrections depending on edge effects \cite{Kamien2023kikibouba}, justifying the inclusion of 90° grain boundaries as generic constituents of the colloidal smectic ground state.

We model these energy-free grain boundaries by extending the layer-based description in \cite{Machon2019}; in addition to \emph{layers} and \emph{half-layers} we introduce \emph{domain walls}, which are also codimension-one objects in both two- and three-dimensions. By construction, a domain wall separates two domains of orthogonal layering. On one side, layers are parallel to the wall, which lies at a minimum-density locus (a half-layer); on the other side, layers meet the wall perpendicularly.
Layer structures that are either perpendicular to the domain wall on both sides or parallel to it on both sides are not topologically protected from locally relaxing to the standard Smectic-A ground state and are therefore not considered. Thus, our complete layer structure is a union of codimension-one submanifolds that may be identified as either layers, half-layers, or domain walls. Defects may reside either strictly on layers or strictly on half-layers, in which case they are classified by the smectic theory in \cite{Machon2019}. However, they may also reside at intersections of half-layers and domain walls, from which maximum-density layers are excluded. In what follows, we assume that defects belong to the latter type.

\begin{figure}[htb]
    \centering
    \includegraphics[width=0.6\linewidth]{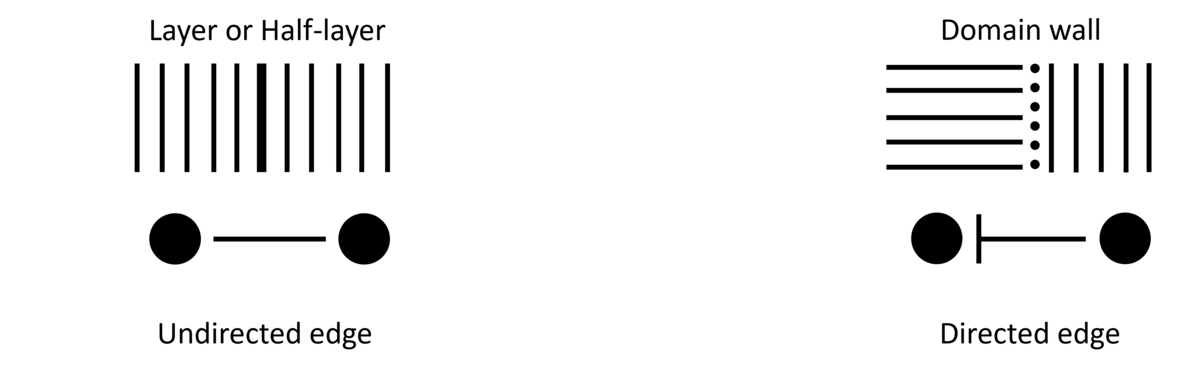}
    \caption{Graph representation of the local layer structure. The intersections between the measuring sphere and the interfaces define the edges of the graph, while the regions between these intersections define its vertices. Left: a layer or half-layer is represented by an undirected edge. Right: a domain wall separating two orthogonal smectic domains is represented by an oriented edge. The side of the \(\vdash\) symbol indicates the side on which the layers are perpendicular to the domain wall. The opposite orientation is represented analogously.  }
    \label{fig:graph_repr}
\end{figure}

To characterize the charge of point defects located at the ends of a grain boundary, we utilize a measuring circle (or sphere in three dimensions) in a manner that is reminiscent of the approach used in the smectic case in \cite{Machon2019}. As in the smectic scenario, we examine how the measuring circle intersects the layer structure, which may now comprise half-layers and domain walls. 
We use a graph representation to describe the structure around defects. The half-layers and domain walls that intersect the measuring manifold split it into domains, represented by the vertices of a graph. The intersections themselves are represented by directed and undirected edges. The direction of the directed edges indicates the side of the domain wall where layers are perpendicular to it (\autoref{fig:graph_repr}).

\section*{Results}

\subsection*{Disclination points in two dimensions}

\begin{figure}[htb]
    \centering
    \includegraphics[width=0.5\linewidth]{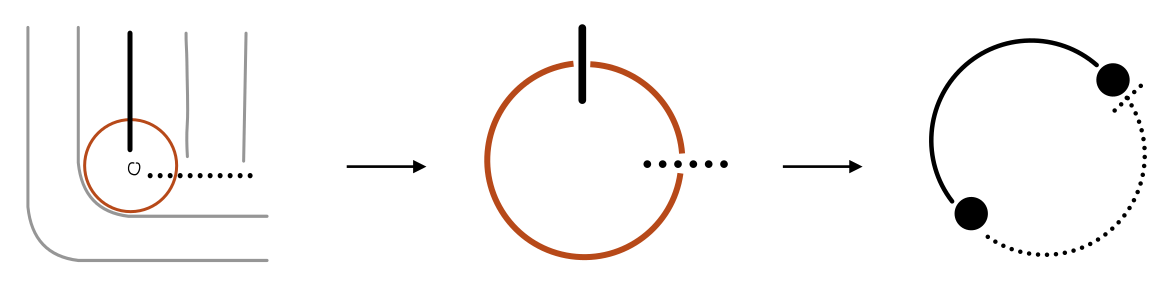}
    \caption{Example of a two-dimensional colloidal-smectic point defect and its graph representation. Reading the interfaces intersecting the measuring circle in cyclic order gives a necklace over three types: half-layer edges and two oriented domain-wall edges. Reflections are not identified, since they generally correspond to distinct layer connectivities. In this example, the disclination charge is \(q= \frac{1}{4}\), given by \autoref{eq:2Dcharge}, as there is one unoriented edge and one oriented edge in the necklace.}
    \label{fig:measuring_circle_necklace}
\end{figure}

In two dimensions, the measuring manifold \(S_\varepsilon\) is a circle, and the associated graph is a cycle graph. The graph edges belong to three types: undirected half-layer edges and two oriented domain-wall edges. Thus, a defect is represented by a necklace over an alphabet of size three: a cyclic word, defined up to rotation, whose letters record the edge types encountered around \(S_\varepsilon\). The length \(n\) of the necklace is the number of intersections between the layer structure and \(S_\varepsilon\) (\autoref{fig:measuring_circle_necklace}). This necklace provides a complete local topological description of the defect. Cyclic relabelings of the necklace are equivalent. However, reflections are not identified, since reversing the cyclic order generally changes the connectivity of the layer structure and cannot be achieved by a smooth deformation without changing the layer topology.

The disclination charge is determined by the total rotation of the director accumulated along a measuring circle \( S_\varepsilon \) traversed counterclockwise around the defect. Along domain walls, the nematic order is discontinuous: the director undergoes a jump of $\tfrac{\pi}{2}$ across an infinitesimal distance. However, the tetratic order remains continuous across the boundary. Consequently, the discontinuity in the tetratic order is limited to isolated points \cite{Wittmann2021particle, Monderkamp2021, Loewen2022, wittmann2024layer}, and thus the winding number, or disclination charge \cite{Mermin1979, Selinger2024}, is well-defined. In regular smectics, this charge can be calculated based on the number of layers intersecting at the disclination point \cite{Poenaru1981, Machon2019}, as previously described. We extend this relation to two-dimensional colloidal smectics by showing that the disclination charge of a point defect can also be inferred from the interface structure meeting at that point. If a defect is associated with \( m \) half-layers (or unoriented edges in the necklace description) and \( r \) domain walls (or oriented edges), the total director rotation around the measuring circle yields the following charge (\autoref{app:charge}), see \autoref{fig:measuring_circle_necklace}:
\begin{equation}\label{eq:2Dcharge}
q = 1 - \frac{m}{2} - \frac{r}{4}
\end{equation}
Consequently, the allowed charges are quarter-integers, and the charge remains a topological invariant satisfying \(q \leq 1\), as in ordinary smectics. In the absence of domain walls, corresponding to the smectic-A phase, the formula reduces to \(q = 1 - \frac{m}{2}\) \cite{Poenaru1981, Machon2019}.
While the relation between layer structure and disclination charge remains informative, q provides only a coarse topological descriptor in the colloidal regime. Neither the disclination charge nor the pair $(m,r)$ constitutes a complete classifier, since distinct, non-equivalent necklaces can share the same values of $q$, $m$, and $r$. A refined classification, therefore, requires the necklace representation, as it captures the specific internal connectivity of domain walls and half-layers that charge alone fails to distinguish.

The topological degeneracy of a specific charge $q$ can be quantified by counting the number of distinct admissible necklace configurations that are associated with $q$. The result is given by $N_q$, the coefficient of the monomial $x^{4(1-q)}$ in the polynomial 
\begin{equation} 
\label{eq:Nx}
N(x)=1+\sum_{n=1}^{4(1-q)}\frac{1}{n}\sum_{d\mid n}\varphi(d)\,\bigl(x^{2d}+2x^d\bigr)^{n/d},
\end{equation}
where $\varphi$ is Euler's totient function. The full derivation of Eq.~\eqref{eq:Nx} is in \autoref{app:combinatorics}. Representative values are provided in \autoref{tab:qC_necklaces_primarycombos}. While the theory permits any negative quarter-integer charge, strong negative charges are physically suppressed; the significant director rotation required by such defects necessitates high gradients that are energetically unfavorable.

\begin{table}[htb]
\centering
\renewcommand{\arraystretch}{1}
\begin{tabular}{c c c}
% \begin{tabular}{c c c >{\centering\arraybackslash}m{0.9\textwidth}}
\hline
$q$ &  $N_q$ & Examples\\ \hline
$1$            & $1$  & \raisebox{-0.3\height}{\includegraphics[height=1.5cm]{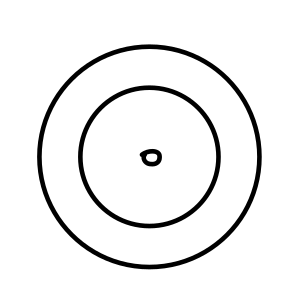}} \\
$\tfrac{3}{4}$ & $2$  & \raisebox{-0.3\height}{\includegraphics[height=1.5cm]{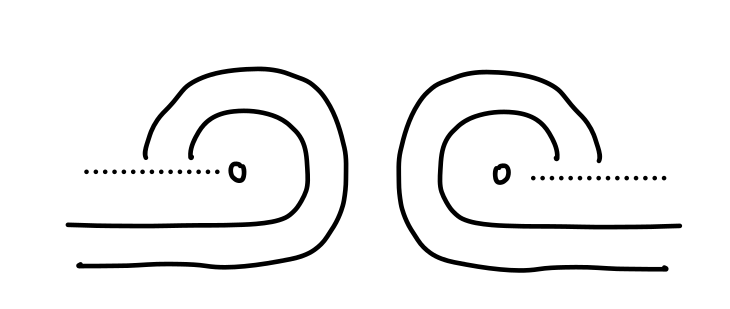}} \\
$\tfrac{1}{2}$ & $4$  & \raisebox{-0.3\height}{\includegraphics[height=1.5cm]{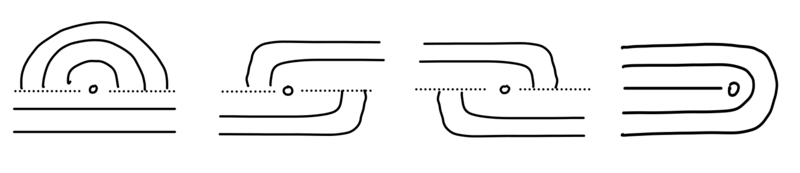}} \\
$\tfrac{1}{4}$ & $6$  & \raisebox{-0.3\height}{\includegraphics[height=1.5cm]{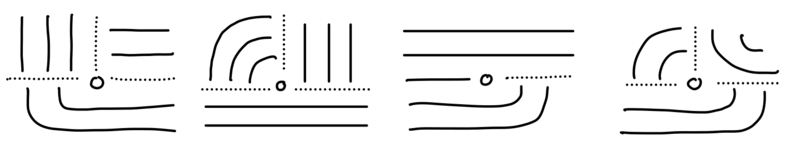}} $ \dots$ \\
$0$            & $11$ & \raisebox{-0.3\height}{\includegraphics[height=1.5cm]{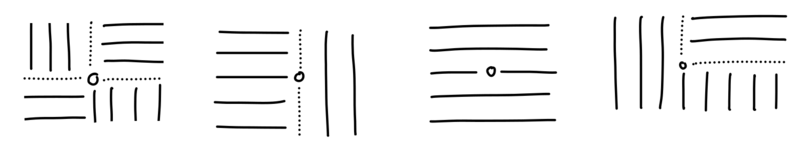}} $ \dots$ \\
$-\tfrac{1}{4}$& $18$ & \raisebox{-0.3\height}{\includegraphics[height=1.5cm]{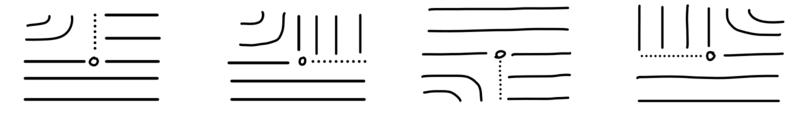}} $ \dots$ \\
$-\tfrac{1}{2}$& $38$ &  \raisebox{-0.3\height}{\includegraphics[height=1.5cm]{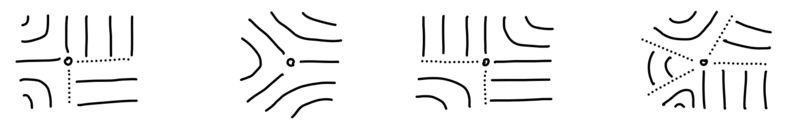}} $ \dots$\\
$\quad \vdots$ & & \\
\hline
\end{tabular}
\caption{Number of topologically distinct defect types $N_q$ for selected charges $q \le 1$, obtained by extracting the coefficient of $x^{4(1-q)}$ in Eq.~\eqref{eq:Nx} (derivation of which is in \autoref{app:combinatorics}).}
\label{tab:qC_necklaces_primarycombos}
\end{table}

Two nearby disclination points can be joined together, and the resulting structure depends on how they are combined. When combining along a common half-layer between two points, or combining transversely to layers, the behavior is quite similar to the combination rules observed in regular smectic-A systems \cite{Machon2019}. This process also incorporates the domain wall interfaces, which are accounted for in the necklace representation.
Moreover, a new type of combination arises in which two disclination points are located at the edges of a single domain wall. In this instance, the combination rules are a blend of the previous cases, where, on one side, it is tangent to a half-layer, and on the other side, it is transverse to layers.
\begin{table}[htb]
\centering
\renewcommand{\arraystretch}{1}
\begin{tabular}{c c c}
\hline
Tangent to layer &  Transverse to layers & Tangent to domain wall\\ 
\hline
\raisebox{-0.3\height}{\includegraphics[height=1.5cm]{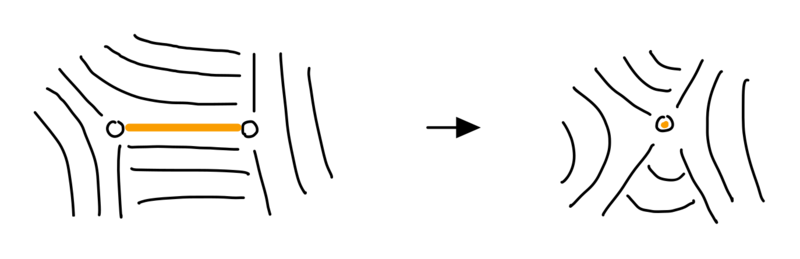}} & \raisebox{-0.3\height}{\includegraphics[height=1.5cm]{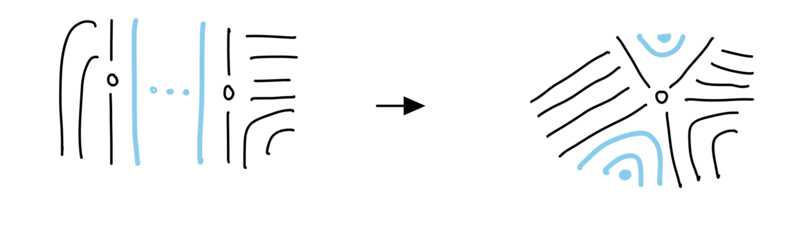}} & \raisebox{-0.3\height}{\includegraphics[height=1.5cm]{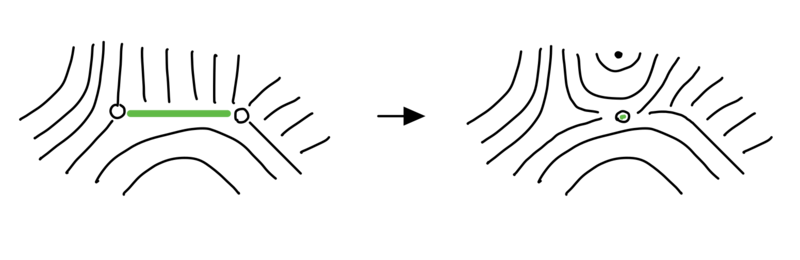}} \\
\raisebox{-0.3\height}{\includegraphics[height=1.5cm]{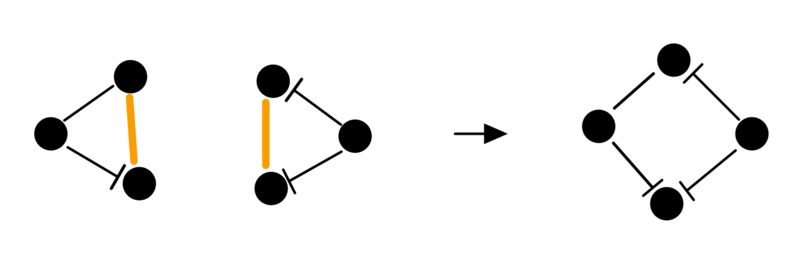}} & \raisebox{-0.3\height}{\includegraphics[height=1.5cm]{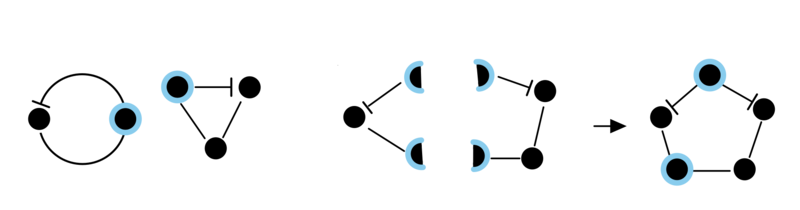}}  & \raisebox{-0.3\height}{\includegraphics[height=1.5cm]{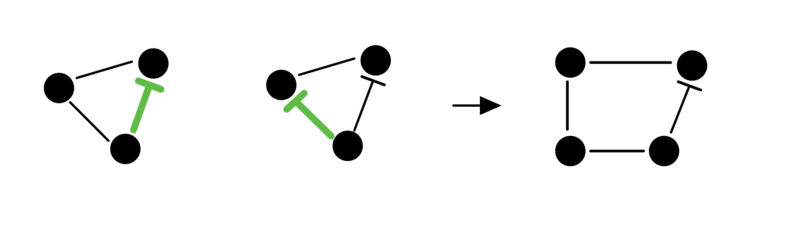}} \\
$q(\mathcal{d}_1\parallel \mathcal{d}_2)= q_1 + q_2 $ & $ q(\mathcal{d}_1 \pitchfork \mathcal{d}_2) = q_1 + q_2 - 1$ &  $q(\mathcal{d}_1 \parallel_{\text{domain wall}} \mathcal{d}_2)=q_1+q_2-\tfrac{1}{2}$ \\
 & two $ +\frac{1}{2}$ elsewhere & one $ +\frac{1}{2}$ elsewhere\\
\hline
\end{tabular}
\caption{Combination rules for two-dimensional colloidal-smectic point defects. The top row shows the combinations, the middle row shows the corresponding necklace operation, and the bottom row gives the resulting local charge. Transverse to layers and tangential to a domain-wall combinations generate additional two or one \(+\tfrac12\) disclinations elsewhere in the texture, respectively.}
\label{tab:combination rules}
\end{table}
The disclination charge of combined point defects can be calculated via the following rules. For combinations that are purely tangential or purely transverse with respect to the layers, the standard smectic-A addition rules remain valid \cite{Machon2019}, with
\begin{equation}\label{eq:combination_rules}
    q(\mathcal{d}_1\parallel \mathcal{d}_2)= q_1 + q_2, \qquad
    q(\mathcal{d}_1 \pitchfork \mathcal{d}_2) = q_1 + q_2 - 1 .
\end{equation}
When two point defects do not share a half-layer or a domain wall, their combination is transverse to the layers (see \autoref{tab:combination rules}). In this case, the local charge of the combined defect is not conserved, and the layers separating the defects generate two additional $+\tfrac{1}{2}$ disclinations. 
Similarly, combining two defects along a domain wall necessarily generates an additional $+\tfrac{1}{2}$ disclination elsewhere in the texture in order to preserve the global layer topology, see \autoref{tab:combination rules}. The effective charge of the combined defect is therefore
\begin{equation}
    q(\mathcal{d}_1 \parallel_{\text{domain wall}} \mathcal{d}_2)=q_1+q_2-\tfrac{1}{2}.
\end{equation}

\subsection*{Disclination lines and points in three dimensions}
In three dimensions, line defects remain locally defined by their transverse structure. Intersecting the texture with a plane perpendicular to the defect line reduces the problem to the two-dimensional classification just described. As a result, the local type of a colloidal smectic line defect is determined by the necklace that describes its transverse slice.
In regular smectics, changes in the transverse structure can occur only at isolated points along the defect line \cite{Machon2019, Aharoni2017}. The layer structure around these ``monopoles'' may be described as augmenting the defect line with cone-like layers that contain one side of the line but not the other. The local transverse data changes discontinuously by an integer at monopoles.
The same structure may appear in colloidal smectics; however, under our ground-state assumptions, a new \emph{domain wall} cannot form such a cone without violating our layer orthogonality assumptions. Thus, monopoles may reside on top of any transverse colloidal smectic structure; however, they only comprise regular smectic layer cones and thus only change the local line charge by a complete integer.

\begin{figure}[htb]
    \centering
    \includegraphics[width=0.5\linewidth]{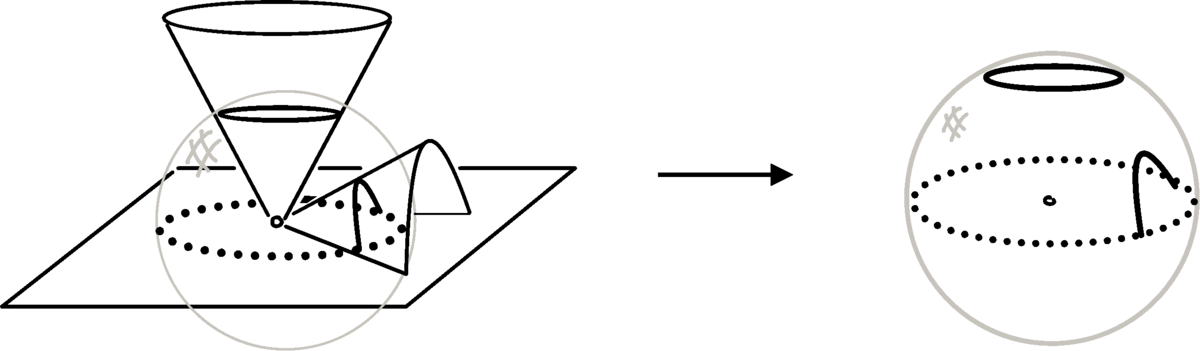}
    \caption{Three-dimensional point defect with a domain wall. The local layer structure is characterized via its intersection with a measuring sphere $S_\varepsilon$. In contrast to ordinary smectics, where the sphere intersects only layer or half-layer curves, the presence of a domain wall introduces additional domain-wall curves on $S_\varepsilon$. Because half-layers may terminate along the domain wall, their intersection curves on the sphere need not close independently, giving rise to arcs whose endpoints lie on the domain-wall curve.}
    \label{fig:3D}
\end{figure}

Point defects are qualitatively different. A defect is represented by a set of curves on the measuring sphere that represent either domain walls or half-layers. Domain walls are represented by closed curves. Half-layers may be either closed or start and end on a domain wall (\autoref{fig:3D}). This curve arrangement, modulo curve homotopies, comprises a full topological classification of point defects in a 3D colloidal smectic. However, this richer structure can no longer be encoded by a tree, as in regular smectics \cite{Machon2019}.

\begin{figure}[htb]
    \centering
    \includegraphics[width=0.5\linewidth]{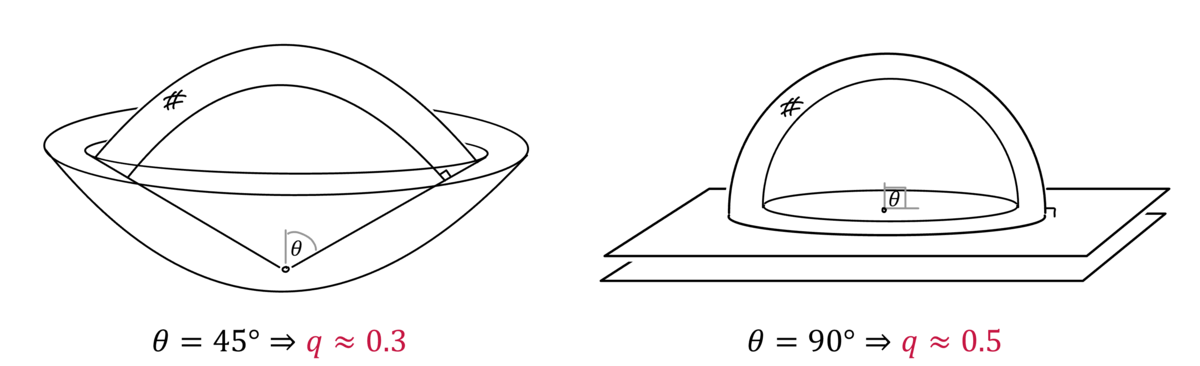}
    \caption{Continuous deformation of a three-dimensional point defect. Varying the cone opening angle \(\theta\) changes the area swept out by the director on the target sphere, and hence changes the hedgehog charge \(q(\theta)=1-\frac{\cos\theta+\sin\theta}{2}\), without creating or annihilating defects.}
    \label{fig:counter_example}
\end{figure}

Moreover, in contrast to two dimensions, the nematic discontinuity across domain walls is not mitigated by a secondary continuous (e.g. tetratic) order. Thus, the disclination ``hedgehog'' charge is no longer a topological invariant. To see this, consider a one-parameter family of configurations in which spherical layers meet conical layers at a right angle, with the cone opening angle $\theta \in (0, \pi/2)$ (\autoref{fig:counter_example}).
The director on the measuring sphere $S^2$ is the outward layer normal; on the upper
spherical-cap portion (polar angle $\theta' \in [0, \theta]$) it points radially,
sweeping area $S_{\rm up} = 2\pi(1 - \cos\theta)$, while on the lower conical portion
($\theta' \in [\theta + \pi/2, \pi]$) it is the outward normal to the cone, sweeping
area $S_{\rm down} = 2\pi(1-\sin\theta)$.
The resulting hedgehog charge is
\begin{equation}
    \label{eq:qtheta}
    q(\theta) = \frac{S_{\rm up} + S_{\rm down}}{4\pi}
    = 1 - \frac{\cos\theta + \sin\theta}{2}.
\end{equation}
Since $q(\theta)$ depends continuously on $\theta$ and is not constant (for instance,
$q(\pi/2) = 1/2$ while $q(\pi/4) \approx 0.29$), the hedgehog charge varies under
a smooth deformation of the layer structure that does not create or annihilate additional defects.

This demonstrates that in a colloidal smectic, the hedgehog charge of a 3D point defect is a \emph{geometric} rather than a \emph{topological} quantity: it depends
on the specific shape of the layer arrangement, and not only on the topological type of the defect, which is given by the spherical curve arrangement. This is in contrast to regular smectics, where the hedgehog charge is an integer (up to a sign) invariant of the underlying nematic order \cite{Mermin1979}, and can be calculated via the defect's tree representation \cite{Machon2019}. 

\section*{Discussion}

We have developed a layer-based framework for classifying topological defects in colloidal smectic liquid crystals, where steric interactions generate orthogonal
domain walls as a generic ground-state feature.
The framework extends the measured-foliation approach of \cite{Poenaru1981, Machon2019} to incorporate domain walls as codimension-one objects alongside layers and half-layers.

A central finding is that the disclination charge carries less topological information than in regular smectics, a result of the introduction of discontinuities in the nematic order. In two dimensions, the charge $q$ is still a genuine topological invariant; it is conserved under continuous deformations of the texture; however, it does not uniquely determine the
defect structure, which is given by a necklace representation. The variety of defects heavily depends on the charge. For instance, there is only one $+1$ defect type, but 151 different $-1$s (see \autoref{app:combinatorics}).
In three dimensions, the disclination charge of point defects in a colloidal smectic is no longer a topological invariant, unlike in a regular smectic, where it is conserved, albeit not being a unique classifier.

The necklace representation of 2D defects is useful not only for classifying individual points but also for illustrating and resolving possible combinations of defects. Two defects may be brought together tangentially to a shared half-layer, transversely to layers, or along a domain wall. In the latter case, the combination is tangential on one side of the wall and transverse on the other. Additional point defects may form during the process, so that the total charge of the combination remains invariant in all cases. The arrangement of layers and additional defects may be extracted from the necklace representation (\autoref{tab:combination rules}).

In this paper, we provided a complete topological classification of 3D point defects by the arrangements of loops and arcs on a sphere. An algebraic or graph-based representation of this classification requires a framework beyond the simple tree graphs used in \cite{Machon2019}. Such a framework may rely on arc complexes \cite{hatcher1991triangulations} or decorated graph representations, and remains an open problem as we have yet to find an elegant one. Such a framework may also naturally capture smectic monopoles \cite{Aharoni2017, Machon2019}, which similarly involve both loops and arcs but allow nontrivial arcs to represent only half-layers and not domain walls. 

\section*{Acknowledgments}
The authors thank Randy Kamien, Jonathan Selinger and Robin Selinger for fruitful discussions. This work is supported by the Minerva Foundation with funding from the Federal German Ministry for Education and Research. C.H. was supported by the Yitzhak Navon Scholarship from the Israeli Ministry of Innovation, Science and Technology.

\bibliographystyle{unsrt}
\bibliography{bibliography}

@book{riordan2014introduction,
  title={An Introduction to Combinatorial Analysis},
  author={Riordan, J.},
  isbn={9781400854332},
  series={Princeton Legacy Library},
  url={https://books.google.co.il/books?id=Sbb_AwAAQBAJ},
  year={2014},
  publisher={Princeton University Press}
}

@article{frenkel1988thermodynamic,
  title={Thermodynamic stability of a smectic phase in a system of hard rods},
  author={Frenkel, D and Lekkerkerker, HNW and Stroobants, A},
  journal={Nature},
  volume={332},
  number={6167},
  pages={822--823},
  year={1988},
  publisher={Nature Publishing Group UK London}
}

@article{Alexander2012,
  author  = {Alexander, Gareth P. and Chen, Bryan Gin-ge and Matsumoto, Elisabetta A. and Kamien, Randall D.},
  title   = {Colloquium: Disclination loops, point defects, and all that in nematic liquid crystals},
  journal = {Rev. Mod. Phys.},
  volume  = {84},
  pages   = {497--514},
  year    = {2012},
  doi     = {10.1103/RevModPhys.84.497}
}

@article{Mermin1979,
  author  = {Mermin, N. David},
  title   = {The topological theory of defects in ordered media},
  journal = {Rev. Mod. Phys.},
  volume  = {51},
  pages   = {591--648},
  year    = {1979},
  doi     = {10.1103/RevModPhys.51.591}
}

@article{Poenaru1981,
  author  = {Po{\'e}naru, V.},
  title   = {Some aspects of the theory of defects of ordered media and gauge fields related to foliations},
  journal = {Commun. Math. Phys.},
  volume  = {80},
  pages   = {127--136},
  year    = {1981}
}

@article{narayan2006nonequilibrium,
  title={Nonequilibrium steady states in a vibrated-rod monolayer: tetratic, nematic, and smecticcorrelations},
  author={Narayan, Vijay and Menon, Narayanan and Ramaswamy, Sriram},
  journal={Journal of Statistical Mechanics: Theory and Experiment},
  volume={2006},
  number={01},
  pages={P01005},
  year={2006},
  publisher={IOP Publishing}
}

@article{Machon2019,
  author  = {Machon, T. and Aharoni, H. and Hu, Y. and Kamien, R. D.},
  title   = {Aspects of defect topology in smectic liquid crystals},
  journal = {Commun. Math. Phys.},
  volume  = {372},
  pages   = {525--547},
  year    = {2019},
  doi = {https://doi.org/10.1007/s00220-019-03366-y} 

}

@article{Aharoni2017,
  author  = {Aharoni, H. and Machon, T. and Kamien, R. D.},
  title   = {Composite dislocations in smectic liquid crystals},
  journal = {Phys. Rev. Lett.},
  volume  = {118},
  pages   = {257801},
  year    = {2017},
  doi     = {10.1103/PhysRevLett.118.257801}
}

@article{geigenfeind2015confinement,
  title={Confinement of two-dimensional rods in slit pores and square cavities},
  author={Geigenfeind, Thomas and Rosenzweig, Sebastian and Schmidt, Matthias and de Las Heras, Daniel},
  journal={The Journal of chemical physics},
  volume={142},
  number={17},
  year={2015},
  publisher={AIP Publishing},
  doi = {https://doi.org/10.1063/1.4919307}
}

@article{Kleman1973,
  author  = {Kl{\'e}man, M.},
  title   = {Defect densities in directional media, mainly liquid crystals},
  journal = {Philosophical Magazine},
  volume  = {27},
  pages   = {1057--1070},
  year    = {1973}
}

@article{Monderkamp2021,
  author  = {Monderkamp, Paul A. and Wittmann, Ren{\'e} and Cortes, Louis B. G. and Aarts, Dirk G. A. L. and Smallenburg, Frank and L{\"o}wen, Hartmut},
  title   = {Topology of Orientational Defects in Confined Smectic Liquid Crystals},
  journal = {Phys. Rev. Lett.},
  volume  = {127},
  pages   = {198001},
  year    = {2021},
  doi     = {10.1103/PhysRevLett.127.198001}
}

@article{wittmann2024layer,
  title={Layer topology of smectic grain boundaries},
  author={Wittmann, Ren{\'e}},
  journal={Liquid Crystals},
  volume={51},
  number={13-14},
  pages={2182--2188},
  year={2024},
  publisher={Taylor \& Francis}
}

@article{ChenAlexanderKamien2009,
  title   = {Symmetry breaking in smectics and surface models of their singularities},
  author  = {Chen, Bryan G.-g. and Alexander, Gareth P. and Kamien, Randall D.},
  journal = {Proceedings of the National Academy of Sciences of the United States of America},
  year    = {2009},
  volume  = {106},
  number  = {37},
  pages   = {15577--15582},
  doi     = {10.1073/pnas.0905242106},
  pmid    = {19717435},
  pmcid   = {PMC2732710},
  url     = {https://www.pnas.org/doi/10.1073/pnas.0905242106}
}

@article{Loewen2022,
  author  = {Monderkamp, Paul A. and Wittmann, Ren{\'e} and te Vrugt, Michael and Voigt, Axel and Wittkowski, Raphael and L{\"o}wen, Hartmut},
  title   = {Topological fine structure of smectic grain boundaries and tetratic disclination lines within three-dimensional smectic liquid crystals},
  journal = {Phys. Chem. Chem. Phys.},
  volume  = {24},
  pages   = {15691--15704},
  year    = {2022},
  doi     = {10.1039/D2CP00060A}
}

@article{Onsager1949,
  title = {The effects of shape on the interaction of colloidal particles},
  author={Onsager, Lars},
  journal={Annals of the New York Academy of Sciences},
  volume={51},
  number={4},
  pages={627--659},
  year={1949},
  publisher={Blackwell Publishing Ltd Oxford, UK}
}

@article{DogicFraden1997,
  author  = {Dogic, Zvonimir and Fraden, Seth},
  title   = {Smectic Phase in a Colloidal Suspension of Semiflexible Virus Particles},
  journal = {Phys. Rev. Lett.},
  volume  = {78},
  pages   = {2417--2420},
  year    = {1997},
  doi     = {10.1103/PhysRevLett.78.2417}
}

@article{Aarts2017,
  author  = {Cortes, Louis B. G. and Gao, Yongxiang and Dullens, Roel P. A. and Aarts, Dirk G. A. L.},
  title   = {Colloidal liquid crystals in square confinement: isotropic, nematic and smectic phases},
  journal = {J. Phys.: Condens. Matter},
  volume  = {29},
  pages   = {064003},
  year    = {2017},
  doi     = {10.1088/1361-648X/29/6/064003}
}

@article{Wensink2013,
  title        = {Phase behaviour of lyotropic liquid crystals in external fields and confinement},
  author       = {Wensink, H. H. and Dussi, S. and Dijkstra, M. and Jackson, G. and Bolhuis, P. G.},
  journal      = {The European Physical Journal Special Topics},
  year         = {2013},
  volume       = {222},
  number       = {11},
  pages        = {3023--3047},
  doi          = {10.1140/epjst/e2013-02075-x},
  url          = {https://doi.org/10.1140/epjst/e2013-02075-x}
}

@article{Wittmann2021particle,
  title        = {Particle-resolved topological defects of smectic colloidal liquid crystals in extreme confinement},
  author       = {Wittmann, Ren{\'e} and Cortes, Louis BG and L{\"o}wen, Hartmut and Aarts, Dirk GAL},
  journal      = {Nature Communications},
  year         = {2021},
  volume       = {12},
  number       = {1},
  pages        = {623},
  doi          = {10.1038/s41467-020-20842-5},
  url          = {https://doi.org/10.1038/s41467-020-20842-5},
  publisher={Nature Publishing Group UK London}
}

@article{Paget2023,
  author  = {Paget, Jack and Mazza, Marco G. and Archer, Andrew J. and Shendruk, Tyler N.},
  title   = {Complex-tensor theory of simple smectics},
  journal = {Nat. Commun.},
  volume  = {14},
  pages   = {1048},
  year    = {2023}
}

@article{hatcher1991triangulations,
  title={On triangulations of surfaces},
  author={Hatcher, Allen},
  journal={Topology and its Applications},
  volume={40},
  number={2},
  pages={189--194},
  year={1991},
  publisher={Elsevier}
}

@article{Kamien2023kikibouba,
  title = {What promotes smectic order: Applying mean-field theory to the ends},
  author = {King, David A. and Kamien, Randall D.},
  journal = {Phys. Rev. E},
  volume = {107},
  issue = {6},
  pages = {064702},
  numpages = {11},
  year = {2023},
  month = {Jun},
  publisher = {American Physical Society},
  doi = {10.1103/PhysRevE.107.064702},
  url = {https://link.aps.org/doi/10.1103/PhysRevE.107.064702}
}

@book{polya2012combinatorial,
  title={Combinatorial enumeration of groups, graphs, and chemical compounds},
  author={Polya, Georg and Read, Ronald C},
  year={2012},
  publisher={Springer Science \& Business Media}
}

@book{Selinger2024,
  author    = {Selinger, Jonathan V.},
  title     = {Introduction to Topological Defects and Solitons},
  series    = {Lecture Notes in Physics},
  volume    = {1032},
  publisher = {Springer},
  year      = {2024}
}

\newpage
\appendix
\section{Calculating the disclination charge in 2D} 
\label{app:charge}

To characterize the charge of the point defects located at the ends of a grain boundary, we use the semi-directed cycle graph representation introduced above. 
Let $\mathcal{d}$ be a point defect where $m$ half-layers and $r$ domain walls meet. Then $S_\varepsilon$ intersects the texture at $m+r$ marked points: $m$ intersections with half-layers and $r$ intersections with domain walls.

Between successive marked points $i$ and $i{+}1$ on $S_\varepsilon$, we decompose the accumulated director rotation into two parts,
\[
\Delta\alpha_{i,i+1}=\alpha_{i,i+1}+\alpha_{i,i+1},
\]
where $\alpha_{i,i+1}$ is the smooth rotation of the director along the arc of $S_\varepsilon$ between the two points, and $\alpha_{i,i+1}$ is an additional contribution imposed by the local interface types at the endpoints (half-layer versus domain wall, and the relevant side of the domain wall).

A half-layer to half-layer transitions contribute $\alpha_{\mathrm{HL\to HL}}=-\pi$, as in the smectic-A case \cite{Machon2019}.
Now consider a neighboring pair consisting of a half-layer intersection followed by a domain-wall intersection. The director associated with a half-layer is tangential to $S_\varepsilon$, whereas the director associated with the perpendicular side of a domain wall is radial. Hence $\alpha_{\mathrm{HL\to DW}}$ must be an odd multiple of $\tfrac{\pi}{2}$,
\[
\alpha_{\mathrm{HL\to DW}}=(2k-1)\frac{\pi}{2},\qquad k\in\mathbb{Z}.
\]
However, the smectic-A structure between the two points rules out all branches except the minimal clockwise rotation,
\[
\alpha_{\mathrm{HL\to DW}}=-\frac{\pi}{2},
\]
and similarly $\alpha_{\mathrm{DW\to HL}}=-\tfrac{\pi}{2}$. Indeed, choosing any other branch would force an additional half-layer attachment at $\mathcal{d}$: the director would become tangent to $S_\varepsilon$ along the arc (away from the wall), implying an extra half-layer intersection with $S_\varepsilon$, or the half-layers would accumulate at $\mathcal{d}$ in a manner analogous to a high-charge nematic defect (e.g.\ a $+2$ winding number), which is forbidden in smectic layer topology \cite{Poenaru1981, Machon2019}. We therefore take the clockwise branch $\alpha_{\mathrm{HL\to DW}}=\alpha_{\mathrm{DW\to HL}}=-\tfrac{\pi}{2}$.
For the same reason, consecutive domain-wall intersections encountered on their perpendicular sides contribute no additional rotation, so $\alpha_{\mathrm{DW\to DW}}=0$.

Summing $\Delta\alpha_{i,i+1}$ over all arcs of $S_\varepsilon$ yields the total accumulated director rotation
\[
\alpha_{\mathrm{tot}} \;=\; 2\pi - m\pi - r\frac{\pi}{2},
\]
and dividing by $2\pi$ gives the winding number (the charge)
\begin{equation}
q=\frac{1}{2\pi}\!\left(2\pi-m\pi-r\frac{\pi}{2}\right)
=1-\frac{m}{2}-\frac{r}{4}.
\end{equation}
Note that $q\le 1$ as in the smectic-A case, but $q\in \tfrac{1}{4}\mathbb{Z}$.

\section{Necklace combinatorics: how many topological defects exist for a given charge?}
\label{app:combinatorics}

We would like to count the number of distinct necklace configurations corresponding to a given charge $q$. Writing $m$ for the number of "$-$" edges, and $r_1$ and $r_2$ for the numbers of "$\vdash$" and "$\dashv$" edges respectively (so that $r=r_1+r_2$), the charge is given by \autoref{eq:2Dcharge}:
\( q = 1-\frac{m}{2}-\frac{r}{4}\).
For our purpose, it would be convenient to define the integer
\[
C_q \coloneqq 4(1-q)=2m+r_1+r_2 \in \mathbb{Z}_{\ge 0}.
\]

Necklace counting is a classical problem in combinatorics \cite{riordan2014introduction}. We count necklaces by their disclination charge using P{\'o}lya's enumeration theorem \cite{polya2012combinatorial} which counts weighed objects by their total weight. Here, $C_q$ is the total weight of the necklace, where we assign weight $2$ to unoriented edges "$-$" and weight $1$ to oriented edges "$\vdash$" or "$\dashv$". By P{\'o}lya, this corresponds to an edge generating function $x^2+2x$, which we substitute into the necklace polynomial \cite{riordan2014introduction}. The number of necklaces with total weight $C_q$ is then given by the coefficient of $x^{C_q}$ in the resulting polynomial, which is:
\begin{equation}\label{eq:generating function for all}
N(x)=1+\sum_{n=1}^{C_q}\frac{1}{n}\sum_{d\mid n}\varphi(d)\,\bigl(x^{2d}+2x^d\bigr)^{n/d},
\end{equation}
where $\varphi$ is Euler's totient function, i.e.\ the number of natural numbers less than $d$ that are relatively prime to $d$ (which is any divisor of $n$).
Note that only necklaces of length $n \le C_q$ contribute to that coefficient, which is why the outer sum ends at $C_q$. For a fixed $q$ (hence fixed $C_q$), the number of necklace configurations of charge $q$ is the coefficient of $x^{C_q}$ in $N(x)$,
\[
N_q = [x^{C_q}]\,N(x).
\]

\end{document}